# The effect of self-motion and room familiarity on sound source localization in virtual environments

Niklas Isserstedt[1], Stephan D. Ewert[1], Virginia Flanagin[2], Steven van de Par[1]

1) Dept. Medical Physics and Acoustics, Cluster of Excellence Hearing4all, Carl-von-Ossietzky Universität, Oldenburg
2) German Center for Vertigo and Balance Disorders, Ludwig-Maximilians-Universität München

**Abstract**

This paper investigates the influence of lateral horizontal self-motion of participants during signal presentation on distance and azimuth perception for frontal sound sources in a rectangular room. Additionally, the effect of deviating room acoustics for a single sound presentation embedded in a sequence of presentations using a baseline room acoustics for familiarization is analyzed. For this purpose, two experiments were conducted using audiovisual virtual reality technology with dynamic head-tracking and real-time auralization over headphones combined with visual rendering of the room using a head-mounted display. Results show an improved distance perception accuracy when participants moved laterally during signal presentation instead of staying at a fixed position, with only head movements allowed. Adaptation to the room acoustics also improves distance perception accuracy. Azimuth perception seems to be independent of lateral movements during signal presentation and could even be negatively influenced by the familiarity of the used room acoustics.

**I. Introduction**

Due to the recent advances in virtual reality (VR) technology, it is possible to recreate auditory sensations which occur in everyday life in a lab setting. This allows to simultaneously study many factors contributing to our ability to localize the position of an auditory event in a room. The position of the auditory event is usually considered to be determined by the availability of acoustic and visual cues (Kolarik et al., 2016). While visual cues dominate over acoustic cues when available and precise (Loomis et al., 1998), Hladek et al. (2013) could demonstrate that auditory distance perception (ADP) is affected by visual cues as well as acoustic cues which could be explained by an optimal sensory integration of visual and acoustic cues weighted by the inverse of the estimated variability in perceived position for each cue (Alais and Burr, 2004). Gil-Carvajal et al. (2016) were able to show that the position of an auditory event in the azimuth plane is independent of visual cues whereas distance perception is dependent on visual cues. It was found that perceived distance of auditory events was biased by visual cues presented in the form of a loudspeaker presented at varying distances from the participant. Part of the participants judged the auralizations as more acoustically distant if the visual distance was increased while others were not affected by visual information (Postma and Katz, 2017).

If no visual cues are available for the localization of an auditory event, localization is solely based on acoustic cues. Acoustic cues can be divided into three subgroups: 1. Cues for directional localization in the horizontal plane (azimuth), 2. Cues for directional localization in the vertical plane (elevation), and 3. Cues for distance perception (Zahorik et al., 2005). The main cues for direction in the horizontal plane are the interaural time differences (ITDs) and interaural level differences (ILDs) which occur due to different pathlengths of the soundwave to both ears (ITDs) or level differences because of the head shadow (ILDs) (Blauert, 1997). In rooms, reverberation influences the interaural correlation of the acoustic signal arriving at both ears which is relevant for ITD-based contributions to directional localization in the horizontal plane (Blauert, 1997). Overall, interaural correlation decreases with increasing signal duration evaluated after onset of the sound presentation when the amount of reverb relative to the amount of direct sound increases over time (Danilenko, 1969). This leads to a broader perceived auditory image which in turn results in a decreased location accuracy (Jeffress et al., 1962). An increase in reverberation level decreases localization accuracy further (Wagener, 1971). Additionally, the "law of the first wavefront" (Cremer and Müller, 1976) improves localization of an auditory signal in the azimuth plane for acoustic signals containing of (multiple) reflections in addition to the direct sound (Cremer and Müller, 1976).

Localization of an auditory event in the vertical plane has been shown to also depend on monaural cues in the head related transfer function (HRTF), which are caused by direction-dependent reflections of the incoming soundwaves on the upper body, the torso, the head and neck as well as the pinna (Batteau, 1967). The HRTF has specific peaks and notches depending on the elevation of the sound source, which can be used for a localization of the sound source in the vertical plane (Batteau, 1967).

Auditory distance perception has been shown to be supported by several monaural and binaural cues. Studies show that perceived loudness inversely correlates to perceived distance between the sound source and the receiver (Petersen, 1990). This perceived distance, however, can be affected by the output level of the sound source leading to a manipulation of the distance perception making the cue less reliable (Blauert, 1997).

The direct-to-reverberant energy ratio (DRR), which describes the ratio of energy present in the direct sound waves compared to the reflected sound waves was found to support distance perception (Bronckhorst and Houtgast, 1999). The amount of energy present in the direct sound decreases with $\frac{1}{r^2}$ leading to a 6 dB decrease in SPL for doubling the sound source distance in a free field condition (Coleman, 1963). The amount of energy present in the reflected diffuse sound in a room is nearly independent of the sound source distance (Coleman, 1963) so that a lower DRR is in principle informative about sound source distance.

Since the specific ratio between direct sound and reverberation depends on the properties of the reverberant environment, distance perception may require familiarity with the room. It has indeed been shown that familiarity with the room acoustics and the used stimuli result in increased accuracy of distance perception compared to unfamiliar room acoustics and stimuli (Mershon et al., 1989). This increase in distance perception accuracy results from the ability to make internal comparisons between the different conditions of already known room acoustics (Philbeck and Mershon, 2002). Brandewie and Zahorik could additionally show that the familiarity of the room acoustics leads to an increase of speech intelligibility, which they explained to be caused by an improved masking of reverb in the brain (Brandewie and Zahorik, 2010). But whether this improved masking can also be used for a more precise distance estimation in room acoustics is not yet evaluated in the literature leaving room to study the effect of familiarity on distance perception in rooms.

For near field ($d$ < 1m) distance estimation, ILD cues play a role as they show significant changes depending on distance for lateral sources while ITD cues largely remain the same (Brungart and Rabinowitz, 1999 or Arend et al., 2016). These findings indicate that head movements can lead to an increase of location accuracy in the horizontal plane (Thurlow and Runge, 1967), reasons being that ITD and ILD cues are ambiguous on a cone centered around the ear (Van Soest, 1929). Interestingly, also horizontal translational movements of the observer were shown to be important for distance perception. Improved distance perception accuracy was shown for distances $d$ > 2m (Ashmead at al., 1995) while accuracy decreased for distances $d$ < 1.5m (Teramoto et al., 2012). Kolarik et al. (2016) introduced the concept of "acoustic tau" (Kolarik et al., 2016) which describes the effects of an acoustic motion parallax: Horizontal movement perpendicular to the source direction leads to changes in intensity, spectrum and binaural cues which can be used for an improved acoustic distance estimation if the velocity of the horizontal movement is known (Kolarik et al., 2016; Genzel et al., 2018).

Many studies focused on studying effects of isolated factors on sound source localization, while only a few studies tried to replicate real-life scenarios by using head rotations or head mounted displays for visual information (e.g., Fichna et al., 2021; Ahrens et al., 2019). Moreover, real-life conditions often include self-motion in enclosed spaces while judging auditory as well as visual distance cues to estimate the position of a sound source. Accordingly, the goal of this study is to combine several factors affecting auditory localization perception in one experiment, to increase the ecological relevance of the tested condition as well as to compare the relative contribution of individual cues' to sound source localization. As typically encountered in realistic conditions (see, e.g., Fichna et al., 2021), sound source localization will include simultaneous estimation of direction and distance, here using a pointing task in a visual virtual environment. Firstly, this study evaluates how localization accuracy of sound sources in VR changes when dynamic self-motion cues as well as visible cues are available in contrast to the availability of only visual or only dynamic cues, or no visual and dynamic cues at all (Experiment 1). In a second experiment, the influence of the familiarity with the room acoustics regarding localization accuracy is evaluated by block-wise placing participants in a baseline acoustic environment where every fifth presentation differs in reverberation time from the baseline. Results should show a significant difference for distance estimation performance with better estimates for presentations with self-motion of the participant and also for familiar room acoustics as distance estimation is largely influenced by DRR which is directly influenced by the room acoustics as well as familiarity (Bronckhorst and Houtgast, 1999; Mershon et al., 1989; Philbeck and Mershon, 2002; Brandewie and Zahorik, 2010; Kolarik et al., 2016; Genzel et al., 2018). Familiar room acoustics and self-motion provide additional binaural cues and DRR samples which should lead to a better distance estimation. Consequently, there could also be an influence regarding azimuth perception, which depends mostly on binaural cues which are influenced to a lesser extent by self-motion or familiarity (Blauert, 1997; Cremer and Müller 1976, Wallach et al., 1949). Results could differ, if a change in room acoustic negatively interferes with the participants' ability to extract the first wavefront (Wallach et al., 1949).

## II. Experiment 1: Effect of self-motion

In the first experiment, the influence of dynamic visual and/or dynamic acoustic cues was investigated by presenting participants a sound from one of 12 different frontal loudspeaker positions indicated by visible loudspeakers in the virtual room. Participants were asked to move horizontally relative to the sound source during some of the presentations. For the horizontal movements, participants stepped to the left and right side which resulted in corresponding simulated shifts in the presented virtual AV environment. In all cases, participants could rotate their head.

## A. METHOD

### Listeners

12 listeners participated in this experiment (age range: 23 –57). All participants had normal stereoscopic vision as well as spatial hearing abilities according to their own statement.

### Apparatus and stimuli

An uncorrelated pink noise burst sequence consisting of 18 bursts, with a duration of $t_{burst} = 30$ ms each, and a total duration of $t = 6$s was presented at different positions inside a virtual room. The signals were generated using the real-time implementation liveRAZR (Schutte, 2021) of the room acoustics simulator (RAZR; Wendt et al, 2014; Kirsch et al., 2023). The room simulation used here includes an image source model and as such is able to accurately vary DRR. In addition, this method was shown to lead to highly plausible and externalized renderings of the simulated sound sources (Stärz et al., 2022) An HTC Vive Pro was used as a head mounted display (HMD) and Sennheiser 650HD headphones were used, calibrated to produce 65 dB SPL for an anechoic source presented at one meter distance. Four different room acoustics conditions consisting of two rooms with different dimensions and two different T60 times were incorporated. The used T60 times and room dimensions are presented in TABLE I.

TABLE I. Room names as well as room dimensions and corresponding T60 times

| Room name | Room dimensions $x, y, z$, (m) | $T_{60}$ (s) |
|---|---|---|
| Room SL | 6, 8, 2.5 (small, S) | 0.6 (low, L) |
| Room SH | 6, 8, 2.5 (small, S) | 1.8 (high, H) |
| Room LL | 10, 16, 4 (large, L) | 0.6 (low, L) |
| Room LH | 10, 16, 4 (large, L) | 1.8 (high, H) |

In each of the four rooms, the position of the sound source is presented with a randomized off-set around its origin point on an 80cm x 100 cm grid (x-direction x y-direction) (see FIG. 1 for origin points and randomization). The range of offsets was 80cm in the horizontal direction, and 100cm in the vertical direction, keeping sources most proximate to their respective origin positions. This prevented listeners from learning the positions of the 12 sources on the grid. Room SL and SH are referred to as the small and Room LL and LH as the big room. Also, the room acoustics in room SL and LL are referred to as room acoustics with low reverb while the room acoustics in room SH and room LH are described as having more reverb. There are 12 different origin points for the sound source differing in x- and y- direction. The used origin points as well as the area in which the actual sound source is positioned are depicted in Fig 1 (origin points: red dots, possible area of sound source presentation depicted with grid).

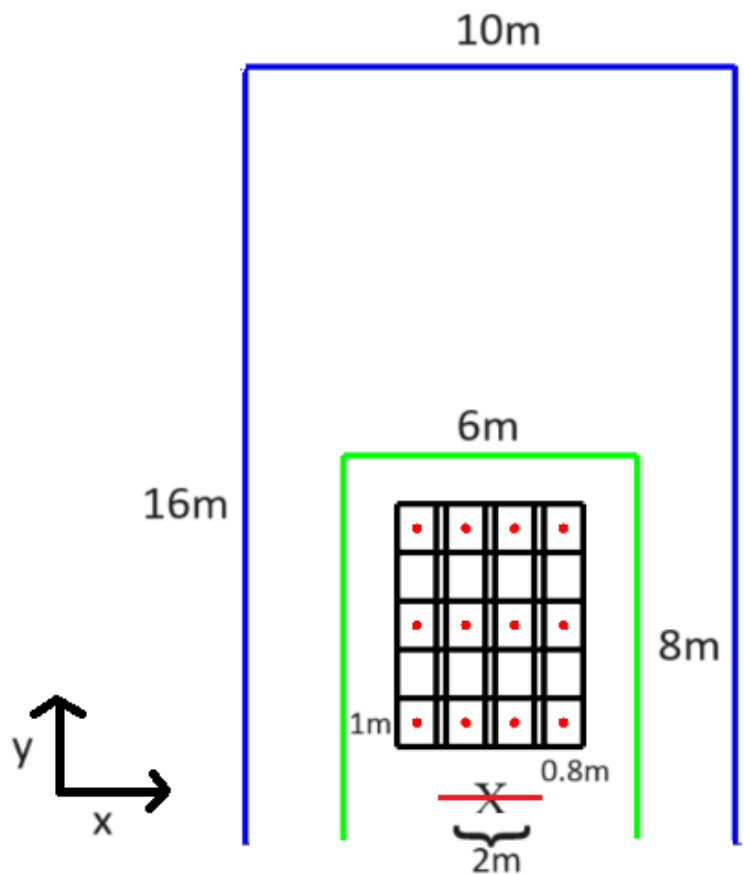

FIG. 1. (Color online). Depiction of the used room dimensions (blue: big room, green: small room), the origin positions of the sound sources (red dots), the grid, in which the sound sources can be placed (black grid), the starting position of the test participant (black X), and the extent to which horizontal movements were allowed (red line); scale is shown.

Additionally, the used rooms are depicted in blue (large room) and green (small room) and the starting position of the test participant is depicted by an X. The extent to which horizontal movements were possible is depicted using a red line.

The areas in which the sound source can be positioned will be called sectors. The sectors are called sector 1 to 12 and are depicted in FIG. 2.

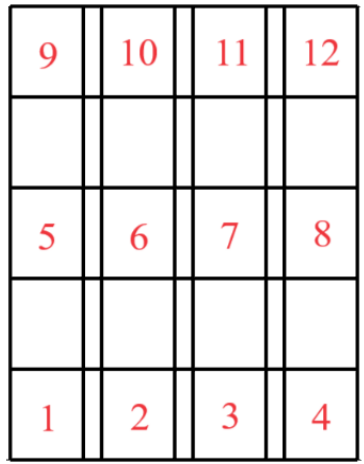

FIG. 2. (Color online). Numbering of the sectors; X represents position of participant, viewing direction in front

Procedure
All participants had to perform all (full factorial) combinations of the four variables used in this experiment resulting in 16 conditions in total. The variables used in this experiment were:

- *Room size* (small, large)
- *Room acoustics* (low and high $T_{60}$ times)
- *Horizontal movement of the test participant* (with, without)
- *Additional visual information* (with and without visual representation of possible source positions by loudspeaker models)

Each of the 12 potential positions of the sound source were presented three times in a block-wise manner for each condition consisting of a selection of the parameters listed above. Each condition was presented twice, while conditions with identical room acoustics were presented after each other to allow the participants to adapt to the current room acoustics.

Before beginning the actual measurement, an information display was presented in VR indicating whether horizontal movement should be performed or not. The area in which the participant could safely move was indicated visually on the virtual floor. Additionally, the starting point of the test participant at the beginning of a condition was marked. The stimulus was only played once. After stimulus presentation, participants indicated the perceived location of the sound source using a hand-held controller with which they pointed a virtual laser beam to the perceived position of the source on the floor. Visual feedback was provided by displaying a loudspeaker model at the indicated position if no additional loudspeaker models were presented or by displaying a black dot at the selected position when 12 loudspeaker models at possible source positions were shown. Positions of the 12 loudspeaker models were chosen by adding an identical offset in x and y-direction to all 12 origin points. Conditions with and without additional visual information are shown in FIG. 3.

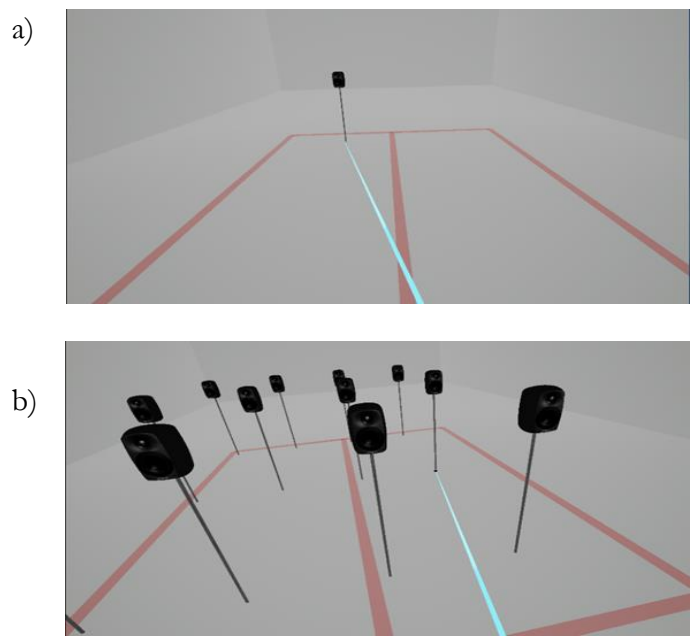

FIG. 3. (Color online). Virtual room displayed via HMD a) without additional visual information, b) with visual information

Participants could change their answer until they pressed a button meant as a confirmation of the chosen position and as a triggering for the next stimulus presentation. Participants were informed, that in the conditions with visible loudspeakers, their locations would not necessarily correspond to the actual position of the sound source. In the actual simulation and rendering one of the loudspeaker models did correspond precisely to the real source position for every trial. A training phase was conducted before the start of the measurements.

**B. Results and discussion**

Since the exact position of the sound source varied for each participant, the difference between the actual position of the sound source and the estimated position of the sound source indicated by the participant was used for further analysis. This difference was broken down into two contributions: 1. The difference in distance between the actual sound source and the estimated position of the sound source as well as: 2. The difference between the azimuth angle between the actual sound source position and the frontal direction (along the y-axis, see FIG. 1) from the position of the participant during the time of the selection of the sound source position, and the according azimuth angle of the estimated sound source.

Since each sound source position varied in one of the 12 sectors, the analysis is made for each sector resulting in 3x12x2 values (3 values for each of the 12 sectors, measured 2 times) per participant and condition for the difference between the actual and perceived sound source distance and azimuth. Following this, outliers were detected and removed using the median absolute deviation (MAD) procedure (Huber, 2004, Rousseeuw and Croux, 1993). A measurement point was deemed an outlier if the MAD of this point was equal or bigger than 3 (Leys et al., 2013). In the next step, the root-mean-square (rms) value for distance und azimuth were calculated for identical source positions for each participant and condition, resulting in 12 values per condition and participant for differences in distance and azimuth. Following this, the obtained data was averaged across participants for each condition resulting in 12 values (one for each sector) for the difference in distance as well as in azimuth for each condition. In total there are 16 x 12 = 192 values for distance and azimuth each. These values were subjected to a repeated-measures analysis of variance (rmANOVA) with four within-subject factors / variables mentioned above. The interactions were assessed too, and the significance level was set to $a = 0.05$. The data used for the rm-ANOVA is shown in TABLE II for distance differences and TABLE III for azimuth differences. Two separate rm-ANOVAs were performed distance and azimuth.Additionally, the mean differences in distance for each of the 12 sectors separated in x- (lateral) and y- (frontal) direction were calculated in the same manner using the mean-value instead of the rms-value as used for absolute errors, to quantify potential biases in the responses of the participants. These data are depicted graphically in FIG. 4.

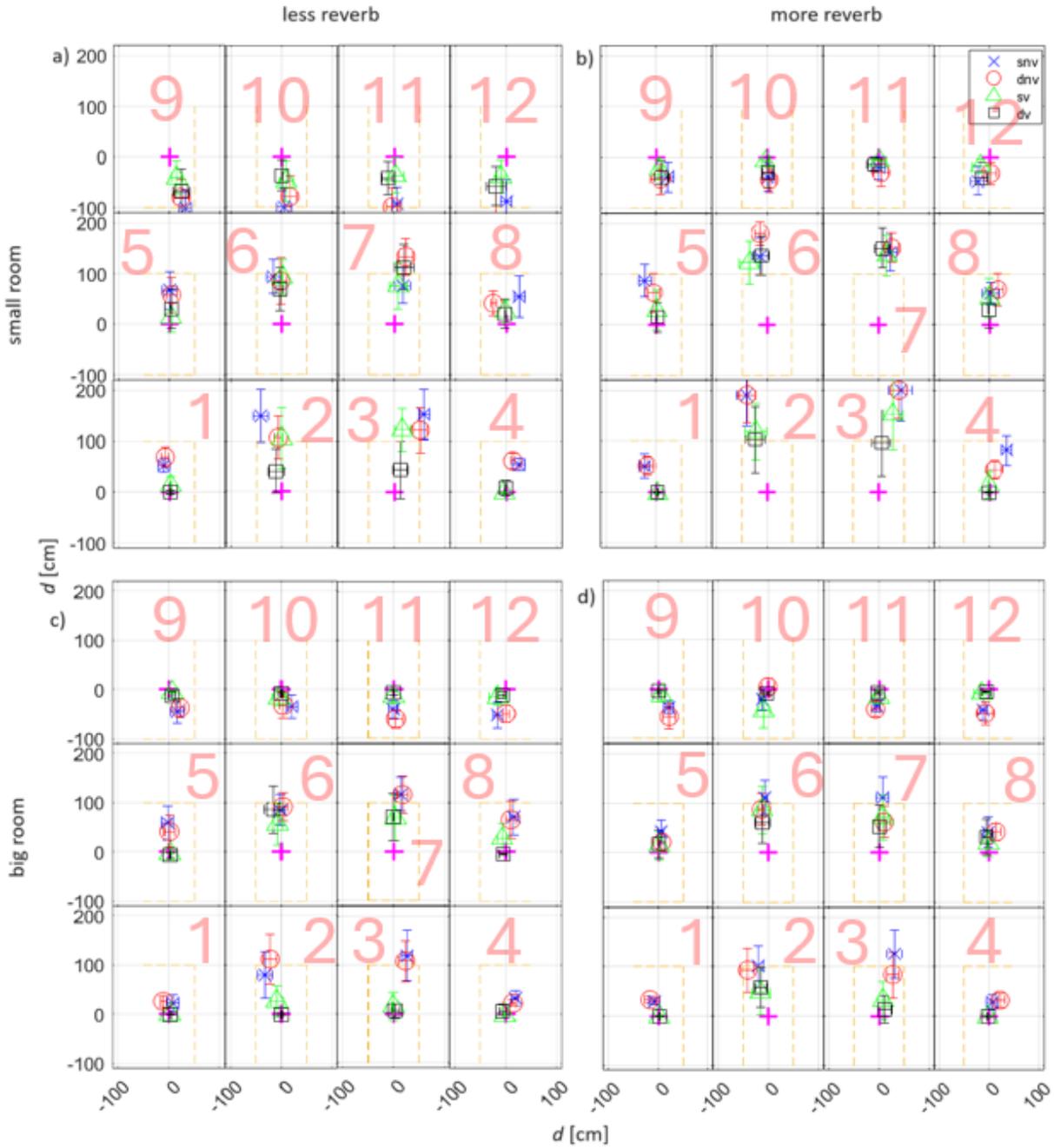

FIG. 4. (Color online). Mean values for perceived location relative to their respect origin points (pink plus signs) depicted in x- as well as y-direction with corresponding standard deviation; for each of the four rooms with a) Room SL, b) Room SH, c) Room LL, d) Room LH. All possible combinations of variables are depicted, blue cross: static and nonvisible condition (snv), red circle: dynamic and nonvisible condition (dnv), green triangle: static and visible condition (sv) and black square: dynamic and visible condition (dv). Yellow dotted lines show the boundaries separating just in between adjacent loudspeakers. Red numbers represent sector number.

FIG. 4 shows mean values for perceived location relative to their respect origin points (pink plus signs) as given by the participants for different combinations of the variables together with their standard deviation for each sector (blue: static and nonvisible condition, red: dynamic and nonvisible condition, green: static and visible condition, black: dynamic and visible condition). Additionally, boundaries signaling where the area of the neighboring loudspeaker begins, are depicted as yellow dotted lines.

A general trend can be observed that deviations from the true source location observed in y-direction seem to be much higher than that in x-direction (up to 10 times higher for y-direction compared to x-direction). Additionally, there does not seem to be any significant bias of the test participant regarding estimations on the x-axis as there is no clear trend towards negative x-values (bias to the left) or positive x-values (bias to the right), which holds true for all sectors and rooms. One can only see a symmetric trend in which positions in the x-direction are slightly overestimated resulting in a higher azimuth angle depending on the location of the sound source. Sources to the left of the listener lead to negative deviation in x-values while sound sources to the right lead to positive values. However, there is a clearer bias in y-direction. Participants tend to overestimate the distance in y-direction for sectors 1-8 while they underestimate the distance in y-direction for sectors 9-12. This is in line with recent literature which states that near distances are overestimated by participants while further distances tend to be underestimated (Zahorik et al., 2005, Parseihian, Jouffrais and Katz, 2014).

This bias tends to be higher for smaller Room dimensions (Room 1 and Room 2) indicating that smaller room dimensions tend to result in a worse localization of sound sources for identical T60-times. Additionally, deviations in y-direction are highest for sectors 2, 3, 6 and 7 in each room which are the sectors located directly in front of the test participant. This leads to cases -especially in Room 1 and Room 2- in which the participants did not locate any or only a few of the sound sources in sectors 2, 3, 6 or 7 but located them one row of sectors behind as most of the values are behind the dotted line indicating the boundaries of the next sector in which a loudspeaker can be placed (sector number 6 instead of number 2, 7 instead of 3, 10 instead of 6 and 11 instead of 7). However, the bias seems to decrease in y-direction for the exterior sectors indicating that it was easier for the participants to estimate the distance of the sound source in y-direction if ITD- and ILD-cues are higher. There are also some cases in which sound sources were located into the area of a neighboring loudspeaker in x-direction (mainly in sectors 2 and 3) indicating again that higher ILD- and ITD-cues lead to more accurate results for localization in the x-direction for near sound sources.

In general, conditions without visual information (blue crosses as well as red circles) tend to differ more from the origin point -indicating perfect localization accuracy- than conditions with visual information (green triangles and black squares) suggesting that visual information helps improve localization accuracy. Comparing static conditions and dynamic conditions (blue crosses and red circles as well as the green triangles and the black squares) with each other, it can be noticed that static conditions (blue crosses and green triangles) differ more from the origin point than dynamic conditions (red circles and black squares) in most of the sectors suggesting that horizontal movement seem to improve localization accuracy in rooms. FIG. 5 depicts the rms-values for each Room and sector (same colors for variable combinations as in FIG. 4; crosses for distance and circles for azimuth). These rms values are the basis for the statistical analysis.

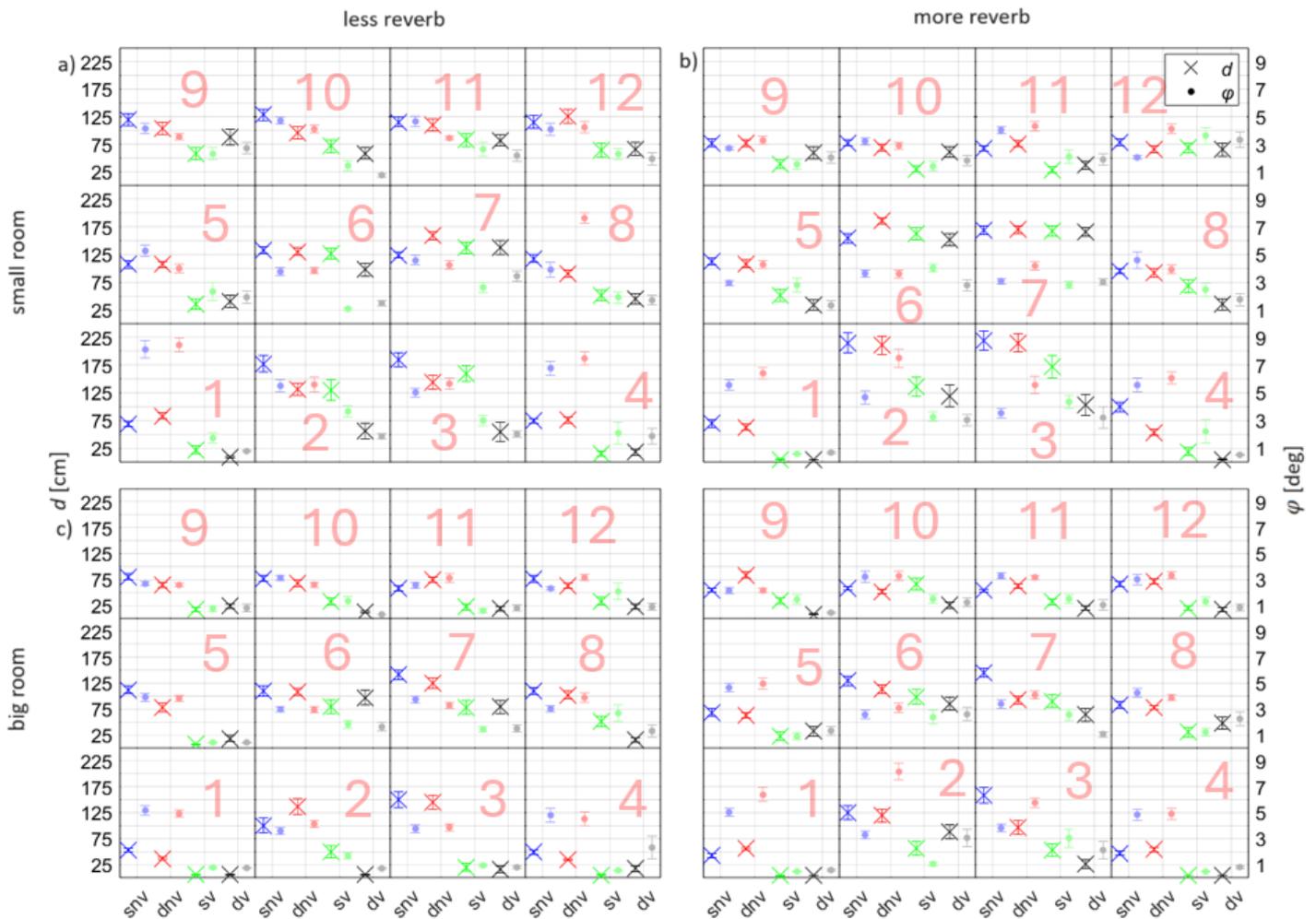

FIG. 5. (Color online). Rms-values for the absolute deviation in distance (crosses) and azimuth (circles) as well as their standard error calculated over all participants for each sector and different combinations of variables with a) Room SL, b) Room SH, c) Room LL, d) Room LH (top left to bottom right); x-axis now shows different conditions: blue: static and nonvisible condition (snv), red: dynamic and nonvisible condition (dnv), green: static and visible condition (sv), black: dynamic and visible condition (dv). Red numbers represent sector number.

FIG. 5 supports the findings stated above. Again, absolute differences in distance estimation are highest for sectors 2, 3, 6 and 7 independent of room acoustics. And again, distance estimation accuracy seems to improve for exterior sectors. For azimuth the opposite phenomenon can be observed. Here the exterior sectors in each room seem to have slightly higher absolute deviations in azimuth.

Again visual information and room size influenced distance and azimuth accuracy. FIG. 5 shows again a better distance as well as azimuth estimation accuracy for larger room acoustics independent of T60-time (Room 3 and Room 4). In general, conditions with visual information (black and green symbols) tend to have the smallest values for both azimuth and distance again suggesting that visual information plays a major role in improving localization accuracy of sound sources if present and coherent as the blue and red symbols are plotted higher in almost every sector in every room. Comparing dynamic conditions without and with visual information (green and black symbols) a small trend can be observed that the dynamic conditions with visual information (black symbols) seem to be plotted below static conditions with visual information (green symbols) both for azimuth and distance suggesting a benefit of horizontal movement for sound localization accuracy if visual information are present. Comparing conditions without visual information (blue and red symbols), one can see a trend that static conditions without visual information (blue symbols) seem to have higher values than dynamic conditions without visual information (red symbols) for absolute differences in distance while no clear trend can be observed in azimuth. This suggests that horizontal movement improves distance perception accuracy while it is unaffected for azimuth perception. But as both static conditions with visual information (green symbols) as well as dynamic conditions with visual information (black symbols) show smaller values than their counterpart conditions in terms of visual information there seems to be an additional benefit of visual information.

The results of the rm-ANOVA support the observations made by analyzing FIG. 4 and 5. The rm-ANOVA showed a significant main effect for room size ($F(1,11) = 38.91$, $p < 0.001$ for distance and $F(1,11) = 126.72$, $p < 0.001$ for azimuth), availability of visual cues for both distance and azimuth perception ($F(1,11) = 71.12$, $p < 0.001$ for distance and $F(1,11) = 43.29$, $p < 0.001$ for azimuth) and availability of horizontal movement for distance perception ($F(1,11) = 6.29$, $p = 0.029$) as can be observed in FIG. 4 and FIG. 5.

This result for distance can be explained by taking into account that the DRR values for larger room sizes will be higher than for small room sizes allowing for a more accurate distance perception (Zahorik et al., 2005). Moreover, due to the geometry of this experiment reflected sound waves will take longer to reach the test participant resulting in a longer time period in which ILD and ITD cues are not distorted by reverb. This leads to a higher interaural correlation of the incoming signal which allows for more precise distance and azimuth estimations (Wagener, 1971). According to the precedence effect, the position of a sound source is based on the first arriving sound wave and its cues (Wallach et al., 1949). These cues are less disturbed for larger room sizes enabling the participant to form a more accurate perception of azimuth (Cremer and Müller, 1976). Moreover, Kopčo and Shinn-Cunningham stated that distance estimations are based on a mapping of a DRR value to a specific distance using only the DRR value of the better ear (Kopčo und Shinn-Cunningham, 2011). This allows for better distance perception in larger rooms due to higher DRR values compared to the small room, where the direct sound may not be perceptible anymore.

These results, showing an improved accuracy in distance perception, are in line with recent studies of Gil-Carvajal et al. (2016) and Postma and Katz (2017) and could be explained by a multimodal integration of the acoustic and visual cues resulting in a better distance as well as azimuth perception (Stein et al., 2014). The same holds true for azimuth perception accuracy which seems to profit from the addition of coherent visual information. It seems to be the case that participants, who were not informed that one of the visual sources coincided with the actual auditory source, still were biased by visual cues and tended to respond more towards positions where visual loudspeakers were presented.

FIG. 5 as well as the results of the rm-ANOVA demonstrate that there is an increased accuracy in distance perception for horizontal movements of the participants in comparison to no movement. In contrast, there is no significant effect for the estimation of azimuth. This significant effect for distance perception can be explained using the approach of Kopčo and Shinn-Cunningham (2011): At the beginning of the signal presentation the participant already hears the stimulus before moving and gets its first cues for distance perception (DRR, intensity, etc.). After movement of the participant, new cues for distance perception in form of different DRR values (also different intensities etc.) will become available. As a consequence, the test participant can integrate this increased number of available cues and also compare different DRR values and their mappings to specific distances to increase their distance perception accuracy (Kopčo and Shinn-Cunningham, 2011). Moreover, the spectrum of the signal changes with movement towards or away from the sound source by changing the amount of energy present in high frequency bands (Little et al., 1992). In return the test participant could associate a specific pair of the sound source and the position of the test participant with a specific distance and could compare further pairs of the same source position but different position of the test participant with this reference pair to increase the accuracy of the distance perception. These explanations are based on the principals of "acoustic tau" (Kolarik et al., 2016) and "motion parallax" (Kolarik et al., 2016) which state an increase for distance perception accuracy for horizontal movements of the test participant (Kolarik et al., 2016, Zahorik et al., 2005). The nonsignificant effect of horizontal movements for azimuth perception can be explained by the generally more accurate perception of azimuth supported by the precedence effect (Cremer and Müller, 1976, Wallach et al. 1949). The horizontal movement of the test participant may not provide much extra information on top of the static azimuth perception also because using motion parallax would require the participant to have an accurate estimate of the extend of horizontal movement made.

Also for distance perception and azimuth perception the interaction between visual information and room acoustics is significant ($F(1,11) = 7.41$, $p = 0.020$ for distance and $F(1,11) = 7.17$, $p = 0.021$ for azimuth) resulting in pairwise comparisons with Bonferroni correction showing an increased accuracy of distance perception if visual cues are present for both room acoustics ($p < 0.001$) while there is no significant effect if the room acoustics are different while the availability of visual information is the same ($p > 0.05$). This result can again be explained using the above-mentioned explanation for the significant effect of the availability of visual cues.

Also there is a significant interaction between room size and room acoustics ($F(1,11) = 16.24$, $p = 0.002$) for azimuth perception with pairwise comparisons with Bonferroni correction showing an increased accuracy of azimuth perception for equal room acoustics if the room size is bigger ($p < 0.001$) which again can be explained using the less disturbed ITD and ILD cues in a bigger room to form a more accurate azimuth perception by the test participant (Wagener, 1971).

Interactions between room size, availability of visual cues and horizontal movement are also significant for azimuth perception ($F(1,11) = 14.74$, $p = 0.003$). Pairwise comparisons with Bonferroni correction show that there is an increased accuracy in azimuth perception for higher T60 times in small room acoustics if no visual information is present. Considering that both critical distances are smaller than the smallest used distance, this could be a hint that participants are able to extract information regarding azimuth using early reflections. This is in line with observations made by Grosse et al (2018) who could show that spatial information can be extracted using direct sound and an early reflection. This information in combination with the information regarding azimuth perception obtained by the direct sound could be used to form a better azimuth perception in this case by using a weighted combination of all information available.

Finally, the interaction between visual cues and horizontal movement is also significant for azimuth perception ($F(1,11) = 11,25$, $p = 0,006$), for which pairwise comparisons with Bonferroni correction show an increased azimuth perception for conditions in which no visual cues are available, and no horizontal movements are permitted in comparison to conditions without visual information but possible horizontal movements ($p = 0,032$). This result could be explained by a negative effect of ITD and ILD cues which are influenced by reverb after movement of the test participant. This influence of reverb leads to a broader sound image which decreases azimuth perception accuracy (Wagener, 1971). However, if visual cues are present there is a significantly better azimuth perception ($p < 0,001$) for conditions in which horizontal movement was permitted in contrast to static conditions. This result indicates that the availability of visual cues outweighs the ILD and ITD cues which are influenced by reverb, because participants can move towards the speaker which they first perceived as playing the signal and then can use the ILD and ITD cues which are obtained after movement to verify their decision if visual cues are present. If no visual cues are present, there is no visual orientation for the participants leading to a decreased performance if horizontal movements are permitted.

### III. Experiment 2: Effect of room familiarity

The second experiment evaluates the influence of the familiarity with the room acoustics on sound source localization. To assess the effect of familiarization, after every four stimulus presentations in one room acoustics, the room acoustics switches for one single presentation, after which it switches back to the first room acoustics. In all conditions the tests participants were instructed to move laterally which has resulted in the most accurate distance perception in Exp. 1.

#### A. Method

Unless otherwise stated, the same listeners, apparatus, rooms, source positions, and stimuli as in Exp. 1 were employed here. Only differences in the methodology are described explicitly.

Procedure

In contrast to Exp 1, the participants were always instructed to move halving the number of conditions with three variables left in this experiment:

- *Room size* (small, large)
- *Room acoustics* (low and high $T_{60}$ times)
- *Additional visual information* (with and without visual representation of possible source positions by loudspeaker models)

In each of the resulting 8 conditions, the effect of familiarity of the room acoustics was assessed by presenting four stimuli after each other using the familiar room acoustics, and then the unfamiliar room acoustics for one stimulus before switching back to the familiar room acoustics. Note that each of these pairs of rooms had the same dimensions but differed only with respect to T60 time. Both possibilities of decreasing or increasing the T60-time were used. Here the sectors for the unfamiliar room acoustics were chosen at random, but all sectors had a chance of roughly 8.5% of being chosen (range from 6.63% to 9.82%).

**B. Results and discussion**

The obtained data was analyzed in the same way as in Exp. 1, including using the MAD to catch outliers. Given that there are now the familiar and unfamiliar room acoustics in each condition, the analysis of the data was split into analyzing deviations in perceived sound locations for the familiar and unfamiliar room acoustics. In contrast to Exp. 1, a mean value for differences in distance and azimuth per participant was now calculated independent of sector for each condition. The reason for that is that due to the random selection of sectors in which unfamiliar stimuli were presented, there may be no presentations of the sound source in some sectors for the unfamiliar room acoustics. Thus, calculating a mean value for each sector is not possible. Following the calculation of the mean values for difference in distance or in azimuth, a comparison between the differences between the familiar as well as unfamiliar room acoustics are possible using an ANOVA with four factors: *Room size*, *Room acoustics*, *Familiarity of the room acoustics* and *Availability of visual cues*. Again, two separate rm-ANOVA were performed to analyze the distance and azimuth differences. The data used for the rm-ANOVA for distance differences is shown in TABLE IV and that for the rm-ANOVA for azimuth differences is shown in TABLE V which can be found in the supplementary material. Here a significance level of $a = 0,01$ was used to minimize the type I error due to different sample sizes for the calculation of values for the familiar vs. the unfamiliar room acoustics.

FIG. 6 shows the mean deviations separated in x- and y-direction using the mean value instead of the rms-value to see potential biases.

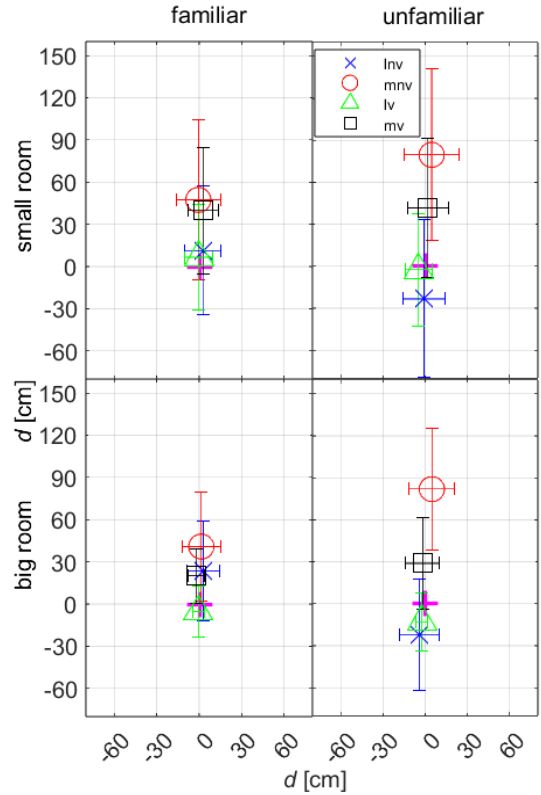

FIG. 6. (Color online). Mean values for deviations in distances separated into differences in x- as well as y-direction as well as their standard deviation; smaller rooms in top row and bigger rooms in bottom row; familiar conditions in first column and unfamiliar in second column; all possible combinations of variables are depicted (blue crosses: less reverb, nonvisual ([nv], red circles: more reverb, nonvisual [mnv], green triangles: less reverb, visual [lv] and black squares: more reverb visual [mv]); origin points marked with pink plus sign

The top two panels show the results for the smaller rooms while the bottom two panels show the same for the larger rooms. The first column shows the results for the familiar room acoustics, the second column shows the results of the unfamiliar one. Different colors were used to depict the different variable conditions.

FIG. 6 lends further support to the results of Experiment 1: deviations in the y-direction are much larger than the deviations in the x-direction. Also, performance is better in conditions in which visual information is present (green and black colored symbols) compared to conditions without visual information (red and blue symbols). In general, participants were able to estimate the x-position of the sound source pretty accurately in this experiment both for the unfamiliar room acoustics as well as the familiar room acoustics (symbols near value 0 for x direction).

Moreover, the interval in which the values are spread in y direction is larger for the unfamiliar room acoustics than for the familiar room acoustics suggesting an increased localization accuracy of sound sources for an adaptation to the room acoustics. In contrast to experiment 1, conditions with a smaller T60-time show less deviations than conditions with a larger T60-time, suggesting an effect of reverb which leads to better sound source localization accuracy for both familiar and unfamiliar room acoustics if less reverb is present. To understand the magnitude of these deviations betterm absolute deviations for distance as well as azimuth perception for different combinations of the variables are plotted in FIG. 7.

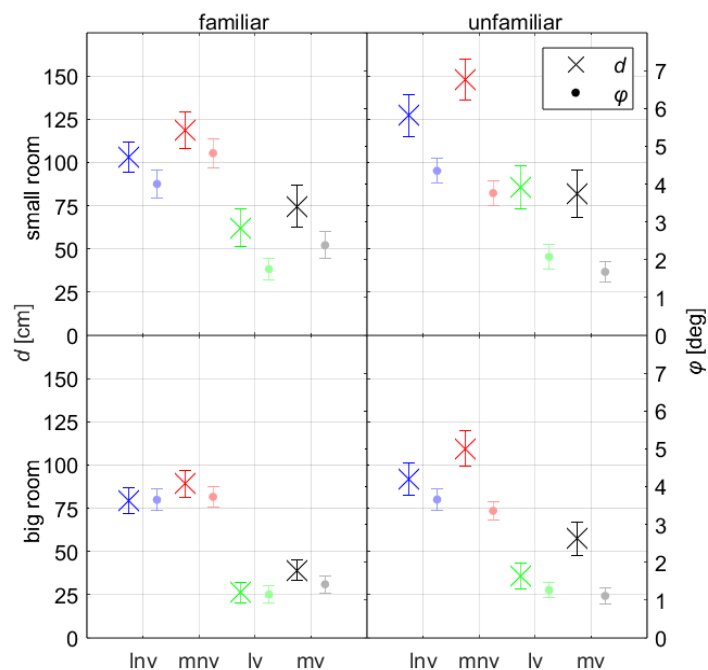

FIG. 7. (Color online). rms values for the absolute deviation in distance (crosses) and azimuth (circles) as well as their standard error calculated over all participants and sectors for the familiar as well as unfamiliar room acoustics; different combinations of variables are displayed: blue: less reverb, nonvisual (lnv), red: more reverb, nonvisual (mnv), green: less reverb, visual (lv), black: more reverb, visual (mv)

FIG:7 shows a tendency for the absolute deviations in distance perception to be smaller in the familiar room acoustics in comparison to estimates made by participants in a unfamiliar room, suggesting an increased distance perception accuracy in a familiar room acoustics. This trend, however, cannot be observed for azimuth perception accuracy. Comparing the amount of reverb for identical room sizes, FIG. 7 shows smaller absolute deviations in distance for smaller amounts of reverb (blue and green symbols) independent of the familiarity of the room acoustics. This could mean a better distance perception accuracy for less reverberant room acoustics independent of familiarity in three of the four plots (only exception: unfamiliar condition in smaller room). However, no difference can be observed for azimuth perception accuracy depending on the T60-time. Also, absolute differences in distance as well as azimuth are smaller for larger room dimensions, again suggesting increased distance as well as azimuth perception accuracy for larger Room sizes.

Additionally, results of the rm-ANOVAs showed that there is a significant effect of Room size [$F(1,11) = 36.70, p < 0.001$ for distance and $F(1,11) = 16.42, p = 0.002$ for azimuth] and availability of visual information ($F(1,11) = 132.04, p < 0.001$ for distance and $F(1,11) = 157.63, p < 0.001$ for azimuth) for both azimuth and distance perception as can be observed in FIG. 6.

Again, distance perception as well azimuth perception accuracy increased with increasing room size as well as in the case in which visual information are present.

Furthermore, the variable Familiarity of the room acoustics proved to be significant for both azimuth and distance perception accuracy ($F(1,11) = 12.28, p = 0.005$ for distance and $F(1,11) = $ 18.09, $p = 0.001$ for azimuth), which is depicted in FIG. 8 and FIG. 9 for differences in x- and y-direction separately.

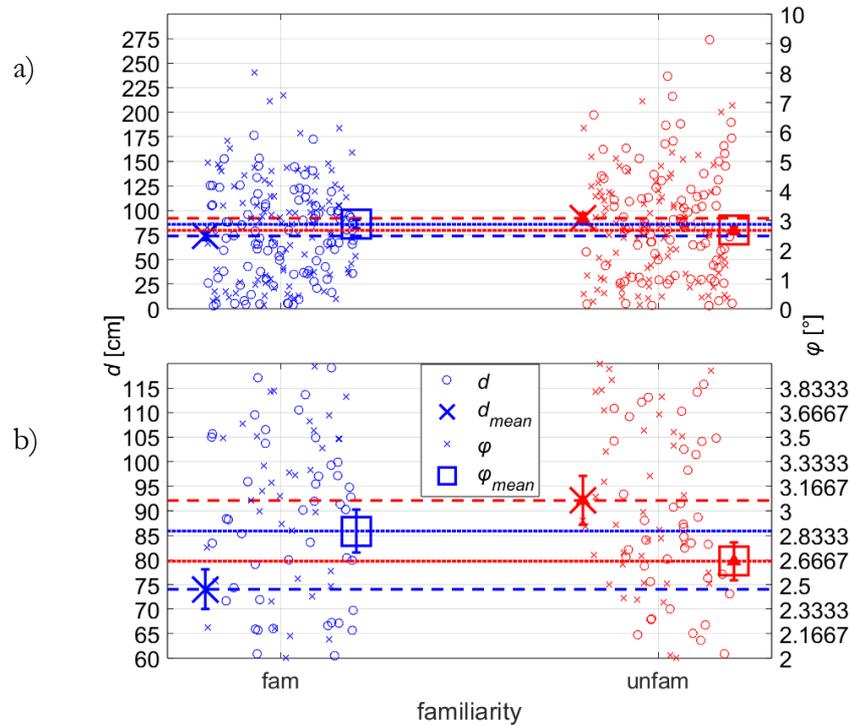

FIG. 8. (Color online). a) Differences in distance (circles) and azimuth (crosses) between sound source position indicated by the test participant and actual sound source position for the two levels of the variable familiarity of room acoustics (fam = familiar, unfam = unfamiliar); additionally, the mean value and standard error for both levels are depicted (cross: mean distance, square: mean azimuth); b) zoomed plots for both distance and azimuth

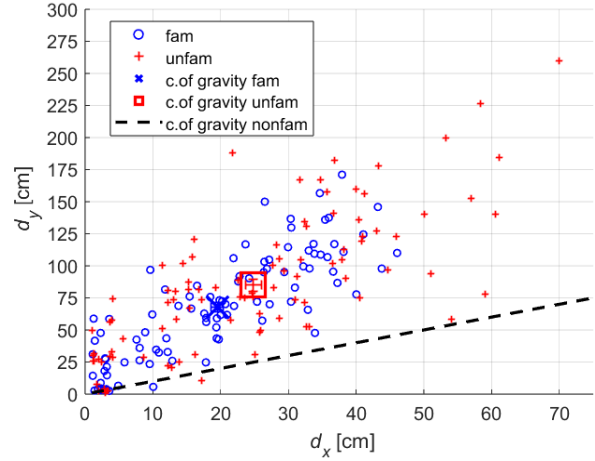

FIG. 9. (Color online). Differences in distance given in difference in x- as well as y-direction for the two different levels of the variable familiarity of room acoustics (fam = familiar, unfam = unfamiliar); additionally, a unity line (dashed black line) and the center of gravity and standard error in x- and y-direction is depicted for each level

Comparisons show that distance perception accuracy is improved when participants could already adapt to the present room acoustics in comparison to a presentation of a stimulus in a unfamiliar room acoustics. This result can be explained by considering that the mapping of cues related to DRR onto distance can only be used in room acoustics where the participants have been able to learn the relation between DRR and distance because this relation will depend on the specific room acoustics. It is not possible to establish such a mapping if the room acoustics is unfamiliar, i.e., when unfamiliar room acoustic was presented much less frequently (1 out of 5 times) compared to the familiar condition.

However, for azimuth perception results show a better perception of azimuth for unfamiliar room acoustics in comparison to room acoustics, depicted in FIG. 8 also. This result is very surprising and may be explained by considering how the azimuth is calculated in this experiment. The difference in azimuth is calculated using the difference in x-positions as well as y-positions of the real position of the sound source, the indicated position of the sound source given by the test participant and the position of the test participant itself. It seems that a familiar room acoustics enables the test participant to judge distance in y-direction more accurately while there is no significant difference in accuracy for an estimation of the position in x-direction. This could lead to significantly worse estimations of the azimuth for familiar room acoustics as the y-direction is estimated more accurately while the x-direction is not. This phenomenon was already explained by taking into account the learning of the DRR to distance mapping in familiar conditions which does not happen in unfamiliar conditions.

A repeated measures ANOVA which divides the distance estimations of the test participant in differences in x- and y-direction shows that the perception of the y-position of the sound source is judged significantly better in familiar room acoustics ($p = 0.002$) while perception on the x-direction is not significantly increased for familiar room acoustics ($p > 0.01$). This result is depicted graphically in FIG. 9 which shows greater differences for mean values including standard errors in y-direction compared to x-direction which in turn suggests that cues that are related to distance perception in y-direction like DRR or intensity profit from familiar room acoustics -likely due to benchmarks for these cues and the availability of comparisons between different cues obtained for the same room acoustics- while ITD and ILD cues are unaffected by the familiarity of the used room acoustics. Here the "law of the first wavefront" (Cremer and Müller, 1976) may dominate in determining the position in x-direction.

Additionally, a significant effect of room acoustics ($F(1,11) = 14.09$, $p = 0.003$) was found for distance perception, while the same variable was insignificant for azimuth perception -suggesting that participants weight cues depending on the amount of reverb (DRR, intensity, etc.) more highly if the room acoustics is switched while azimuth perception is likely independent on room acoustics as it follows the precedence effect (Wallach et al., 1949).

This result also shows that internal weighting of cues can be changed in different circumstances as results for distance perception in Exp. 1 in which the room acoustics was also familiar but not changed during measurement, are insignificant for room acoustics ($F(1,11) = 0,51$, $p > 0,01$) while they are significant in Exp. 2 in which the room acoustics is changed during listening while both experiments used the same room acoustics and room sizes.

**V. General Discussion and Conclusion**

This paper investigated the influence of horizontal movements of the participant during the presentation of a sound source in virtual reality and the influence of familiarity of the room acoustics on distance as well as azimuth perception. Both experiments were conducted using visual VR-technology as well as liveRAZR (Schutte, 2021) for acoustic simulations and used various conditions that differed in room size, availability of visual cues, room acoustics, availability of horizontal movements and familiarity of the room acoustics. Results in general showed a much lower accuracy for distance perception compared to azimuth perception. Two of the three distance ranges where sources could appear could hardly be separated by the participants, manifesting in an overlap of perceived distances for the largest and medium distance range (3m to 6m). Additionally, an increased accuracy for distance as well as azimuth perception for larger room sizes as well as when visual information was present could be shown. Distance perception was significantly improved if participants moved horizontally during presentation of the sound. No significant effect for horizontal movement was observed for azimuth perception. For distance perception, horizontal movements may have provided parallax cues, allowing to convert the perceived azimuth changes into source distance. Horizontal movements might also allow for better distance perception because of an increased number of cue samples regarding DRR, intensity and spectrum of the sound source which might be internally integrated to form a distance perception with increased accuracy (Kolarik et al., 2016, Zahorik et al., 2005, Kopčo and Shinn-Cunningham, 2011). Conversely, azimuth perception was probably dominated by the already accurate static perception resulting from the precedence effect (Cremer and Müller, 1976, Wallach et al., 1949). Since using parallax cues for azimuth perception require an accurate estimate of the extend of horizontal self-movement, the additional information obtained in this manner may have suffered from poor estimates of horizontal self-movements compared to the high acuity with which azimuth can already be perceived in static conditions.

An adaptation to the room acoustics increased distance perception accuracy while it seemed to have no strong influence on azimuth perception accuracy, suggesting that cues for distance perception are weighted more strongly if the room acoustics is already familiar. In this case, benchmarks for cues already exist enabling an internal comparison, while azimuth perception seems largely independent of familiarity of room acoustics. This is supported by the results showing that perception accuracy in y-direction was significantly increased for familiar room acoustics while this was not the case for results in the x-direction. The underlying reason is most likely a mapping of DRR onto distance for familiar room acoustics, which is not available for unfamiliar room acoustics, suggesting that previous exposure to the room acoustics is required for more accurate distance judgements. Interestingly, a change from a familiar room acoustics with less reverb to an unfamiliar room acoustics with more reverb leads to a bias towards increased distance judgements which in line with using the internal 'DRR-to-distance mapping' of the familiar room. It also is in line with recent studies which showed an increase in distance perception if the DRR cue becomes smaller in environments with more reverb (Zahorik et al., 2005). However, no bias towards decreased distance estimates can be found if the familiar room acoustics has more reverb and participants are asked to judge the distance of an unfamiliar room acoustics with less reverb. This finding could be a base for further investigation.

Overall, the findings suggest that participants are able to integrate information obtained by horizontal movements and familiarity of room acoustics to improve distance perception while azimuth perception seems largely independent of horizontal movements and the familiarity of room acoustics. These results suggest that self-motion of participants improves mapping of DRR values, which are strongly dependent on room acoustics and familiarity and therefore increases distance perception accuracy. On the other hand, azimuth perception accuracy is hardly influenced by self-motion or familiarity as binaural cues, which are largely independent on reverberation time (at least at the first wavefront) and self-motion as well as familiarity are used for azimuth perception. Results of this study are particularly interesting for everyday situations in which distance estimation of auditory sources plays a major role (i.e., indoor sport events) or when visual cues are less reliable than acoustic cues (i.e., nighttime situations). The current findings provide insights on how perceptual distance as well as azimuth estimates of sound sources are obtained -and improved- by allowing for self-motion and/or accommodation to unknown room acoustics.

Future research can be based on the results and could focus on finding models for predicting distance perception.

**AKNOWLEDGEMENTS**

The authors thank Henning Hoppe, Stefan Fichna, and Henning Steffens for support with the original experimental setup. This work was supported by the Deutsche Forschungsgemeinschaft, DFG SPP Auditive – Project-ID 444827755.

**AUTHOR DECLARATIONS**

**Conflict of Interest**

The authors have no conflicts to disclose.

**Ethics Approval**

All participants were financially compensated, participated voluntarily, and provided informed consent. The study was approved by the Ethics committee of the University of Oldenburg.

**DATA AVAILABILITY**

The data are available from the authors on request.